\documentclass[page-classic]{epl2}

\title{Relation Between First Arrival Time and Permeability in 
Self-Affine Fractures with Areas in Contact}
\shorttitle{First Arrival Time} 

\author{Laurent Talon\inst{1} \and Harold Auradou,\inst{1} \and
Alex Hansen\inst{2,1}}
\shortauthor{L.\ Talon \etal}

\institute{
\inst{1} Univ.\ Pierre et Marie Curie-Paris6, Univ.\ Paris-Sud, CNRS,
Lab.\ FAST, B{\^a}t.\ 502, Campus Univ., Orsay, F--91405, France.\\

\inst{2} Department of Physics, Norwegian University of Science and
Technology, N--7491 Trondheim, Norway\\
}
\pacs{83.80.Hj}{Suspensions, dispersions, pastes, slurries, colloids}
\pacs{46.50.+a}{Fracture mechanics, fatigue and cracks}
\pacs{62.20.mt}{Cracks}

\abstract{
We demonstrate that the first arrival time in dispersive processes in 
self-affine fractures are governed by the same length scale characterizing
the fractures as that which controls their permeability.  In 
one-dimensional
channel flow this length scale is the aperture of the bottle neck, i.e., the
region having the smallest aperture.  In two dimensions, the concept of a
bottle neck is generalized to that of a minimal path normal to the flow.  
The length scale is then the average aperture along this path.  There is a
linear relationship between the first arrival time and this length scale, 
even when there is strong overlap between the fracture surfaces creating
areas with zero permeability.  We express the first arrival time directly 
in terms of the permeability.
}
\begin{document}
\maketitle

Due to their role in the flow properties of tight and low permeability 
reservoirs such as shale gas reservoirs and carbonate reservoirs, and on
contaminant transport e.g.\ in connection with waste storage, the study of
transport in fractures is still a very vigorous field 
\cite{abmwan91,n96,h06,w09}.  Most present theoretical efforts attempts 
to relate the transport properties of fractures to the statistics of 
the aperture fields through analytical models based on statistical averages, 
weak disorder perturbation expansions \cite{krk95}, mean-field approximations 
or simplified aperture models \cite{bf08}. We also mention the work of
Zhan and Yortsos \cite{zy00} where a method to deduce the heterogenities of 
a permeability distribution from the concentration arrival time field was 
proposed.  

Due to the surface roughness, i.e., the heterogenities of the aperture 
field, these relations provide satisfactory results only over a finite 
range of conditions and do not permit to predict the behavior of a fracture 
with large heterogenities in aperture field. One of the main difficulties 
is to correctly take into account the increasing influence of the contact 
area as the fracture aperture is decreased \cite{mcsg06,wht08,wht09,nwht09}. 
We analyze in this Letter the dispersion problem at finite P{\'e}clet number 
and identify the proper aperture measure for this problem, taking into account 
severe heterogeneities such as large contact zones. Our main focus is on the 
breakthrough time, i.e., the time at which the tracer appears at a given
position.  The surprising result that we find is that this aperture is the 
same as the one controlling the permeability \cite{tah10a,tah10b}.   

There are now numerous experimental studies and field observations that 
demonstrate that natural fractures have a self-affine 
roughness \cite{ptbbs87,blp90,psj92,mhhr92,sgr93} --- for a review, see
Bonamy and Bouchaud \cite{bb11}.  Self-affine fractures are characterized by 
a scaling invariance of the statistical properties of the surface 
roughness under a rescaling of the distances by a factor $\lambda$ in the 
average fracture plane and a rescaling $\lambda^\zeta$ of the heights.
Here $\zeta$ is the Hurst or roughness exponent which takes value close to 
$0.8$ for rocks such as granite \cite{as02} and values close to $0.5$ for 
porous rocks like sandstone \cite{bah98,pavh06}.

We consider in this work synthetic self-affine fracture surfaces that have been
generated using a Fourier method \cite{t97,s98}.  
The fracture is modeled by matching the 
fracture surface with an opposite flat surface.    
Since we use the Reynolds
approximation, this correctly models flow in fractures as it is only the 
aperure that enters the flow equations.   We define the fracture 
aperture $H(\vec r)$ as the distance between the two surfaces at position 
$\vec r$.  In the present work, the rough surface progressively
approaches the flat surface one and the aperture of overlapping regions is 
set to zero. Hence, $H(\vec r) > 0$ where there is no overlap and $H(\vec r)=0$
where there is overlap.   

The flow field is determined for a fixed pressure difference 
between the fracture inlet and outlet by solving the finite differenced 
Reynolds equations through LU decomposition. 
The total volume entering  
the fracture per width and time at the inlet is proportional to the 
pressure 
$\Delta P$ over the fracture, and is given by 
\begin{equation}  
\label{darcy}
Q=-\frac{K}{L\mu} \Delta P\;,
\end{equation} 
where $K$ is the permeability and $\mu$ the viscosity of the fluid.  $L$
is the length of the fracture.  

\begin{figure}
\onefigure[scale=0.6]{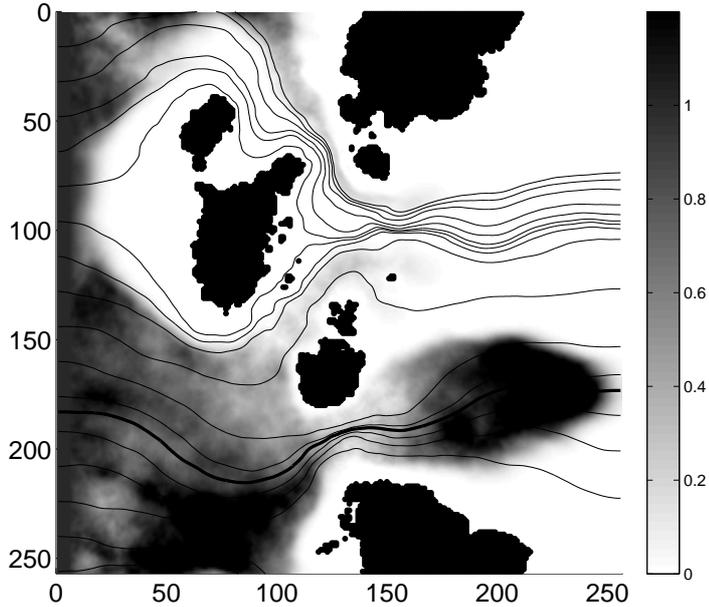}
\caption{
Fluid flows in a self-affine aperture field seen from above.  The curves
are the streamlines of the flow field.  The streamlines have been found using
the Kirchhoff method.  We have then used the Lattice Boltzmann method to
simulate dispersion fixing the P{\'e}clet number at 10.  The tracer 
concentration is shown on a grey scale where darker means higher 
concentration.  Areas where the aperture is zero --- 
i.e., the fracture surfaces
are in contact --- are shown as black. }   
\label{fig:streamlines}
\end{figure}

The breakthrough time is typically obtained by summing along each streamline 
the local time of convection \cite{mtthn88,krb99,mkk06}.  This method does,
however, not take into account diffusion between and along the streamlines.
We have instead discretized the
velocity field on a square lattice with nearest-neighbor and next 
nearest-neighbor connections.  If we assume that the dispersion time
between neighboring nodes is the inverse of the velocity component along
the vector between them, we have then introduced diffusion into the 
dispersion process.  This is so since a given tracer particle will not follow
the streamlines but move between the nodes via projected velocity vectors.
We add to this description the analysis of Stern \cite{s75} of the 
first arrival time of a diffusive process with or against a convective 
velocity field, making it possible to tune the P{\'e}clet number.  
In order to find the first arrival 
time, we use the optimal path algorithm of Hansen and Kertesz \cite{hk04}.  
We have verified our algorithm by comparing it to a two-dimensional 
Lattice Boltzmann 
method \cite{tmrsy03}. Fig.\ \ref{fig:streamlines} shows the tracer 
concentration in gray levels at breakthrough with P{\'e}clet number $Pe=10$ 
based on the Lattice Boltzmann method. In the following, we do not discuss
any dependence of our results on the P{\'e}clet number.  The 
numerical experiments we report have been done at $Pe=10$ as a reasonable
value.  Other values add nothing significant.

Transport properties of fractures are often characterized by equivalent or 
apparent apertures --- such as the hydraulic, mass balance or electrical 
apertures --- which refers  to the aperture of a fracture with flat and 
parallel walls having the same property as the original fracture. 
In practice, equivalent apertures are estimated from hydraulic and 
conservative tracer tests. The mass balance aperture $b_m$ is defined as the 
ratio between the fluid flux $Q$ and the averaged fluid velocity 
$\overline{u}$ \cite{t92,gc00}. 
In practice, the average fluid velocity $\overline{u}$ equals the average of 
the velocities of all fluid particles and should be derived from the average 
residence time determined from the momentents of the time distribution of the 
measured tracer breakthrough curve \cite{t92}. Here we connect the first 
arrival time, $\tau_{min}$, to the average fluid velocity $\overline{u}$ 
by the expression $\overline{u}=L/\tau_{min}$, as proposed by Guimer{\`a}  
and Carrera \cite{gc00}. Since the pressure gradient is kept 
fixed, the mass balance aperture $b_m=Q/\overline{u}$ is proportional 
to $K\tau_{min}$.

Before considering two-dimensional fractures, i.e., fractures
where the aperture is orthogonal to a two-dimensional fracture plane, 
we consider a  {\it one-dimensional\/} version of the problem, i.e., a
fracture where the aperture is orthogonal to a fracture line.
We introduce a cartesian coordinate system with the $x$ axis
along the line which now consitutes the flat surface.  
Let us set $a = \min_x H(x)$. We 
then define 
\begin{equation}
\label{hx}
h(x)=H(x)-a\;.
\end{equation} 
When $a \le 0$, the fracture is closed and hence
the permeability is zero. For positive $a$, it is this parameter that
controls the permeability of the fracture in the lubrication approximation
\cite{tah10b}.  The permeability is in this limit given by the expression
\begin{equation}
\label{1dperm}
\frac{L}{K} = \int_0^L \frac{d\xi}{k(c\xi^\zeta+a)^3}\;,
\end{equation}
where $k$ and $c$ are two parameters. 
$c$, the topothesis which characterizes 
the roughness of the aperture field, is a length scale,
and $k$ has the units of permeability.  For large $a$, this gives rise to
the scaling relation
\begin{equation}
\label{1dpermlargea}
K \sim L a^3\;,
\end{equation}
whereas for intermediate $a$, we find
\begin{equation}
\label{1dpermintera}
K \sim L a^{3-1/\zeta}\;.
\end{equation}
For small $a$, the permeability is completely controlled by the region
where $h(x)=0$, and the continuum approach behind eq.\ (\ref{1dperm})
breaks down.  We then find that the permeability is given by 
\begin{equation}
\label{1dpermsmalla}
K \sim L^0 a^3\;.
\end{equation}

We now calculate $\tau_{\min}$ using the lubrication approximation and in the 
infinite P{\'e}clet number limit where diffusion is absent.  The first tracer 
to traverse the rough channel is the one which has traveled along the 
streamline located midway between the walls {\it i.e.} where the velocity is
at its maximum.  The time this has taken is $\tau_{min}$, and it is given by
\begin{equation}
\label{1arrival1d}
\tau_{\min} = \int_0^L \frac{dx}{u(x)}\;,
\end{equation}
where $u(x)$ is the maximal velocity at position $x$ along the channel 
which is proportional to the flow rate over the local aperture in the 
lubrication aproximation. We have thus
\begin{equation}
\label{1arrival1dbis}
\tau_{\min} \propto \frac{1}{Q} \int_0^L \left(h(x)+a\right)dx\;.
\end{equation}
Combining eqs.(\ref{darcy}) and (\ref{1arrival1dbis}), we get
\begin{equation}
\label{tauk1d}
\tau_{\min} K = \frac{L\mu}{\Delta P} \int_0^L \left(h(x)+a\right)dx\;.
\end{equation}
This integral may be performed by using order statistics \cite{tah10b}: 
We order the function $h(x)\to h[\xi]=h(x[\xi])$ such that  
$h[\xi_1]\le h[\xi_2]$ when $\xi_1\le\xi_2$. 
For a self-affine profile we have that $h[\xi]\sim \xi^\zeta$, and eq.\ 
(\ref{1arrival1dbis}) becomes
\begin{equation}
\label{orderint1d}
\tau_{\min} K 
= \frac{L\mu}{\Delta P} \int_0^L \left( c \xi^\zeta + a\right) d\xi
=\frac{\mu}{\Delta P}\left[cL^{2+\zeta}+ a L^2\right]\;,
\end{equation}
where $c$ is a constant.  Hence, we have the central result for a 
one-dimensional channel
\begin{equation}
\label{central1d}
\tau_{\min} K= A + C a\;,
\end{equation}
where $A\propto L^{2+\zeta}$ and $C\propto L^2$.  Hence, we see that it is
the minimum aperture $a$ which controls the first arrival time $\tau_{\min}$.
This is the same aperture that controls the permeability, see eqs.\
(\ref{1dpermlargea}) -- (\ref{1dpermsmalla}). This is a somewhat
surprising result, since at the minimal apperture location, because of 
mass conservation, the flow rate is maximal. Consequently, the time in 
the bottle neck effect has a small contribution to the 
integral eq.\ (\ref{1arrival1d}). However, the bottle neck controls the
total flow rate, and this, in turn, controls the first arrival time.

As described in Talon et al.\ \cite{tah10a,tah10b}, the extrapolation of 
the bottle neck effect to a two-dimensional fracture is not straight 
forward since the minimal aperture point is easily bypassed by the flow.
However, it is possible to generalize the concept to two dimensions 
by replacing the minimal aperture $a$ by the 
{\it minimal path aperture.\/} In order to introduce this concept, we 
orient our fracture such that the one of the edges parallel to the average
flow direction follows the $x$ axis. The $y$ axis follows the edge where
the tracer is injected and the $z$ axis is orthogonal to the average fracture
plane.  Hence, $0\le x \le L$ and $0\le y \le W$.  
We define ${\cal C}(x)$ as a path starting at $(x,y=0)$ and ending at 
$(x',y=W)$ without crossing itself.  Hence, we define the quantity
\begin{equation}
\label{bx2d}
B(x)=\frac{1}{WL} \left[\min_{{\cal C}(x)} 
\int_{{\cal C}(x)} d\vec{\ell} \cdot \vec{e_y}(\vec{\ell})^3\right]^{1/3}\;.
\end{equation}
This is the minimal average fracture opening over all paths starting at
$(x,0)$ and ending anywhere along the opposite edge at $y=W$.  This
quantity corresponds to $H(x)$ in the one-dimensional case.  The 
{\it minimal path aperture\/} is defined as
\begin{equation} 
\label{bc}
b_c=\min_x B(x)\;,
\end{equation}
corresponding to the smallest aperture $a$ in the one-dimensional case.
We finally define
\begin{equation}
\label{bx}
b(x)=B(x)-b_c\;,
\end{equation}
in the same way we defined $h(x)$ in eq.\ (\ref{hx}) in the one-dimensional
case. 

The central idea in Talon et al.\ \cite{tah10b} was that the two
definitions $b_c$ and $b(x)$ could replace $a$ and $h(x)$ in the 
one-dimensional case in the permeability integral (\ref{1dperm}).  After
ordering $b(x)\to b[\xi]$, we find that $b[\xi] \sim \xi^\beta$, 
where $\beta=1.5$ for $\zeta=0.8$ and $\beta=1.2$ for $\zeta=0.3$. 
The intermediate scaling regime (\ref{1dpermintera}) then is replaced by
\begin{equation}
\label{1dperminterb}
K \sim WL b_c^{3-1/\beta}\;,
\end{equation}
whereas the large and small scale regimes become respectively
\begin{equation}
\label{1dpermlargeb}
K \sim WL b_c^3\;,
\end{equation}
and 
\begin{equation}
\label{1dpermsmallb}
K \sim WL^0 b_c^3\;.
\end{equation}
Numerical experiments based on solving the Kirchhoff equations give
\begin{equation}
\label{2dperm}
K \sim \left\{
       \begin{array}{ll}
              WL b_c^{2.25\pm 0.02} & \mbox{for $\zeta=0.8$}\;,\\
              WL b_c^{2.16\pm 0.02} & \mbox{for $\zeta=0.3$}\;,\\
       \end{array}
       \right.
\end{equation}
for the intermediate regime.  The results are very close to the 
prediction of eq.\ (\ref{1dperminterb}).

\begin{figure}
\onefigure[scale=0.3]{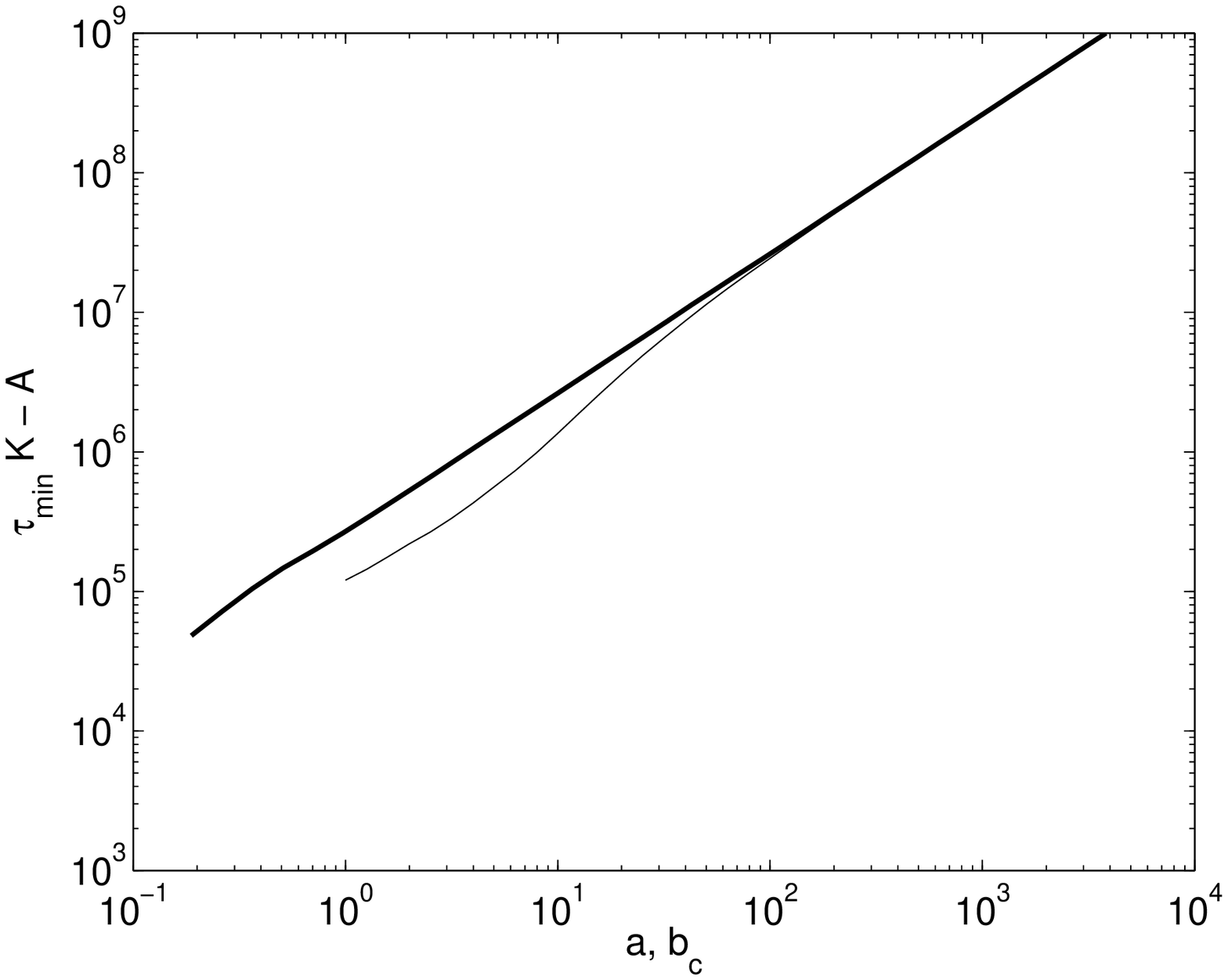} 
\onefigure[scale=0.3]{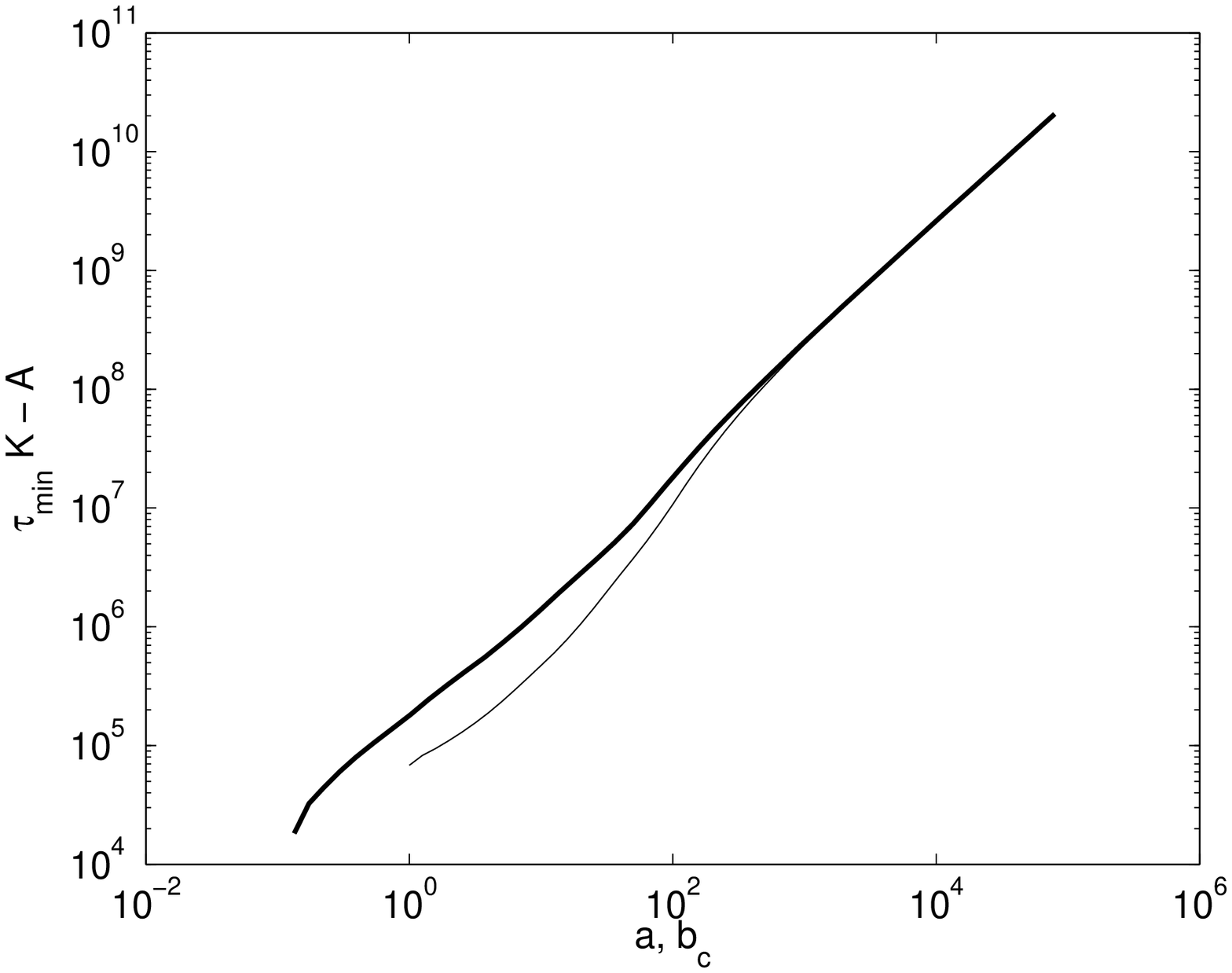} 
\caption{
Loglog plot of $\tau_{min}K-A$ as 
function of $a$ and $b_c$ for $\zeta=0.3$ (upper) and $\zeta=0.8$ (lower). 
The solid curves are for $\tau_{\min}K-A$ vs.\ $b_c$, whereas the 
broken curve is for $\tau_{\min}K-A$ vs.\ $a$.  We determine $A$ by
varying it until we obtain the best possible power law.  In both figures,
the straight portions of the curves have unit slope as indicated in eq.\
(\ref{1arrival2d}).  The curves are based on one sample of size 
$512\times 512$ for each roughness.} 
\label{fig:tau_K_bc}
\end{figure}

\begin{figure}
 \onefigure[scale=0.3]{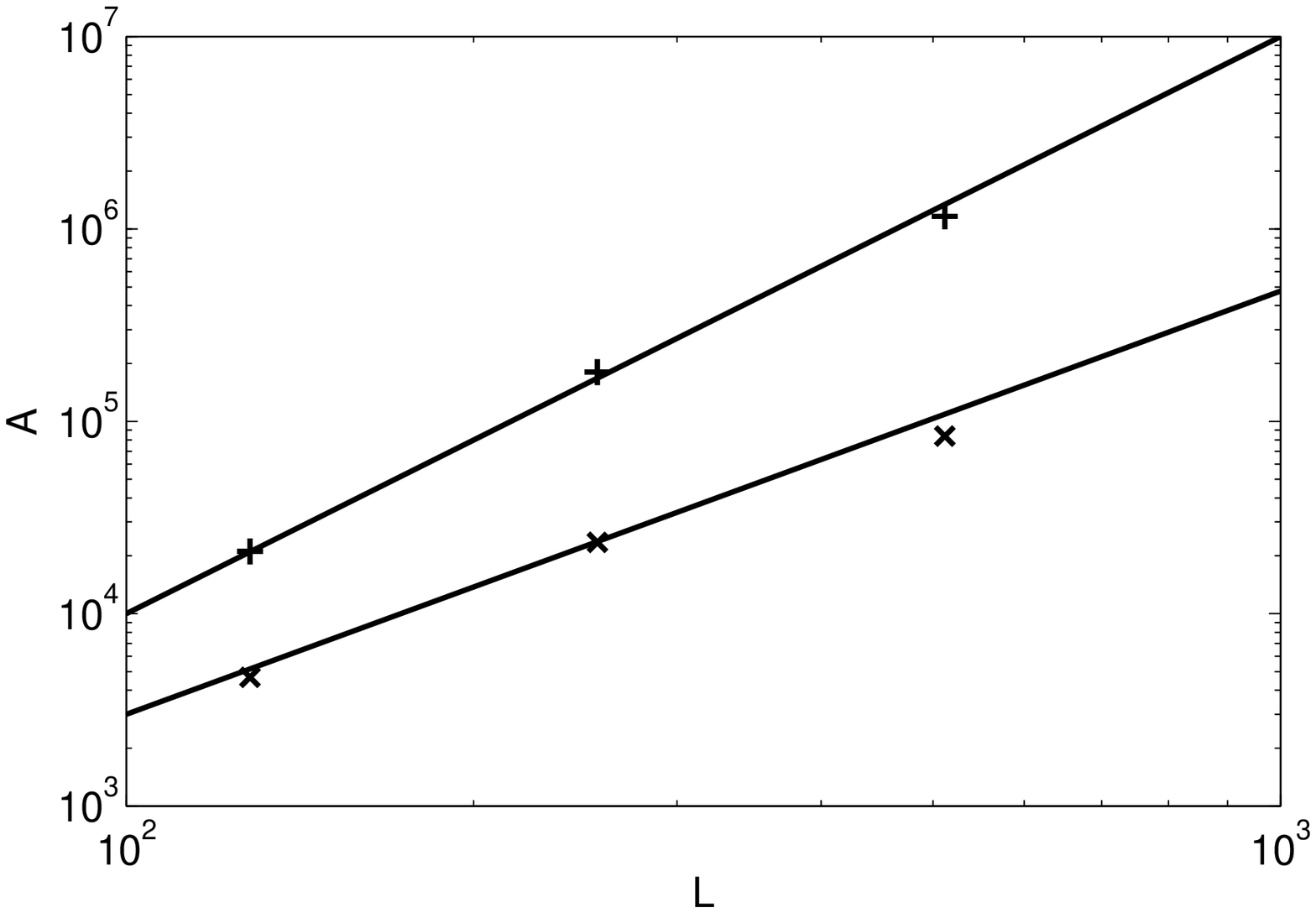}
 \onefigure[scale=0.3]{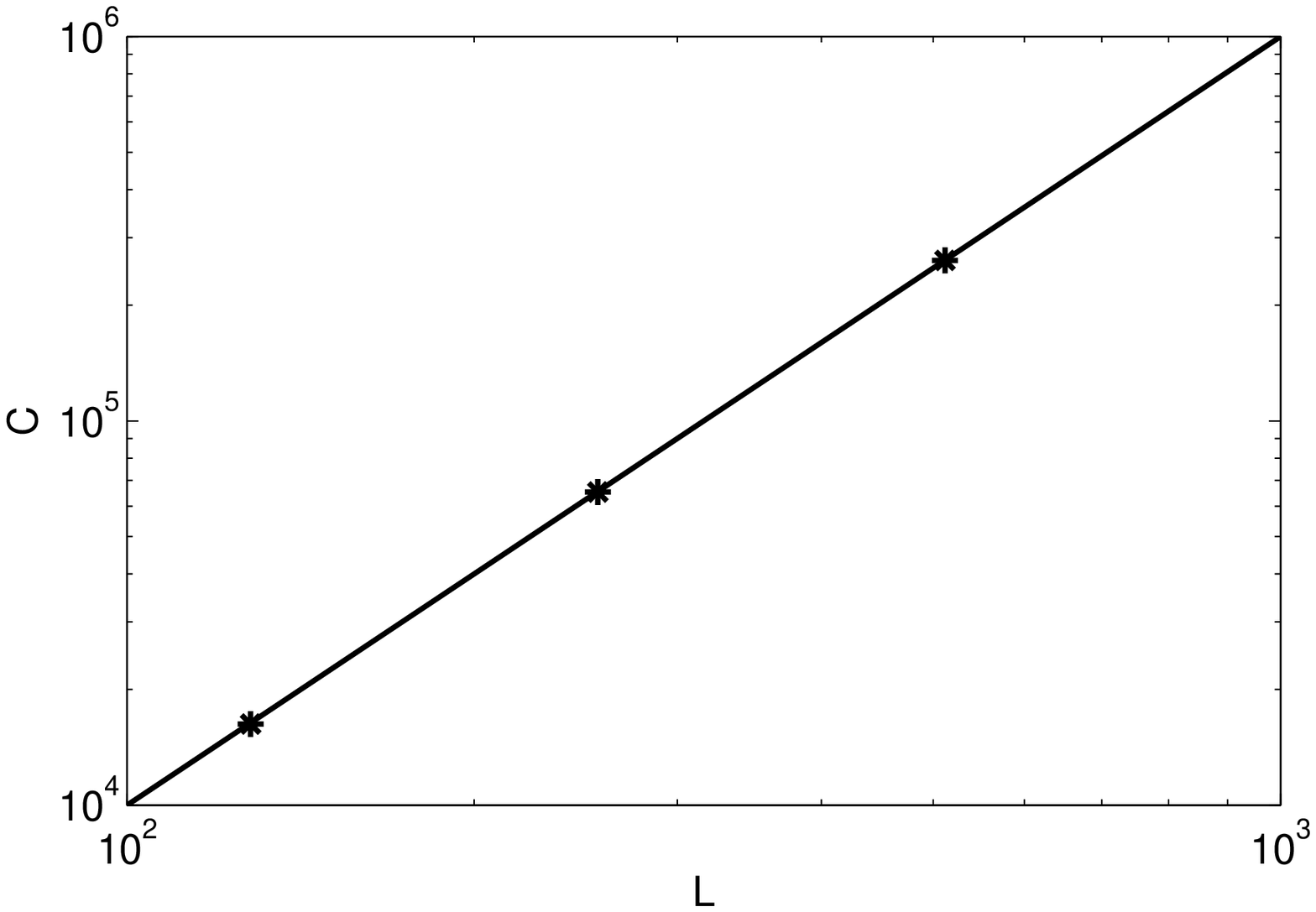}
\caption{Scaling of coefficients $A$ (upper figure) and $C$ (lower figure)
defined in eq.\ (\ref{1arrival2d}).  The straight lines are best fits and
have slopes 3 for the $\zeta=0.8$ data ($+$) and 2.2 for the $\zeta=0.3$ 
data ($\times$) in
the upper figure.  The theoretical values are 2.8 and 2.3 respectively. In
the lower figure, the best fit has slope 2.0 for both the $\zeta=0.8$ and
$\zeta=0.3$ data. The theoretical value is 2.}
\label{fig:scalingAC}
\end{figure}

By following exactly the same procedure for the first arrival time, i.e.,
replace $h(x)$ by $b(x)$ and $a$ by $b_c$ in eq.\ (\ref{1arrival1dbis}), we
find
\begin{equation}
\label{1arrival2d}
\tau_{\min} K = A + C b_c\;,
\end{equation} where $A \propto W L^{2+\zeta}$ and $C \propto W L ^2$.
Fig.\ \ref{fig:scalingAC} shows $A$ and $C$ as a function of $L$ verifying
these two scaling laws.
We show in fig.\ \ref{fig:tau_K_bc}, $\tau_{\min} K$ as a function of $b_c$
and of $a$.  We see that the linearity of $\tau_{\min} K$ is verified for
the entire range of $b_c$ values, whereas it is only true for large values
of $a$. This is where $a$ and $b_c$ begin to coincide.

Hence, we have verified that the miminal path aperture $b_c$ controls
{\it both\/} the permeability $K$ and the minimal time $\tau_{\min}$. This,
together with eq.\ (\ref{1arrival2d}) constitute two main results of this
Letter.

\begin{figure}
\onefigure[scale=0.3]{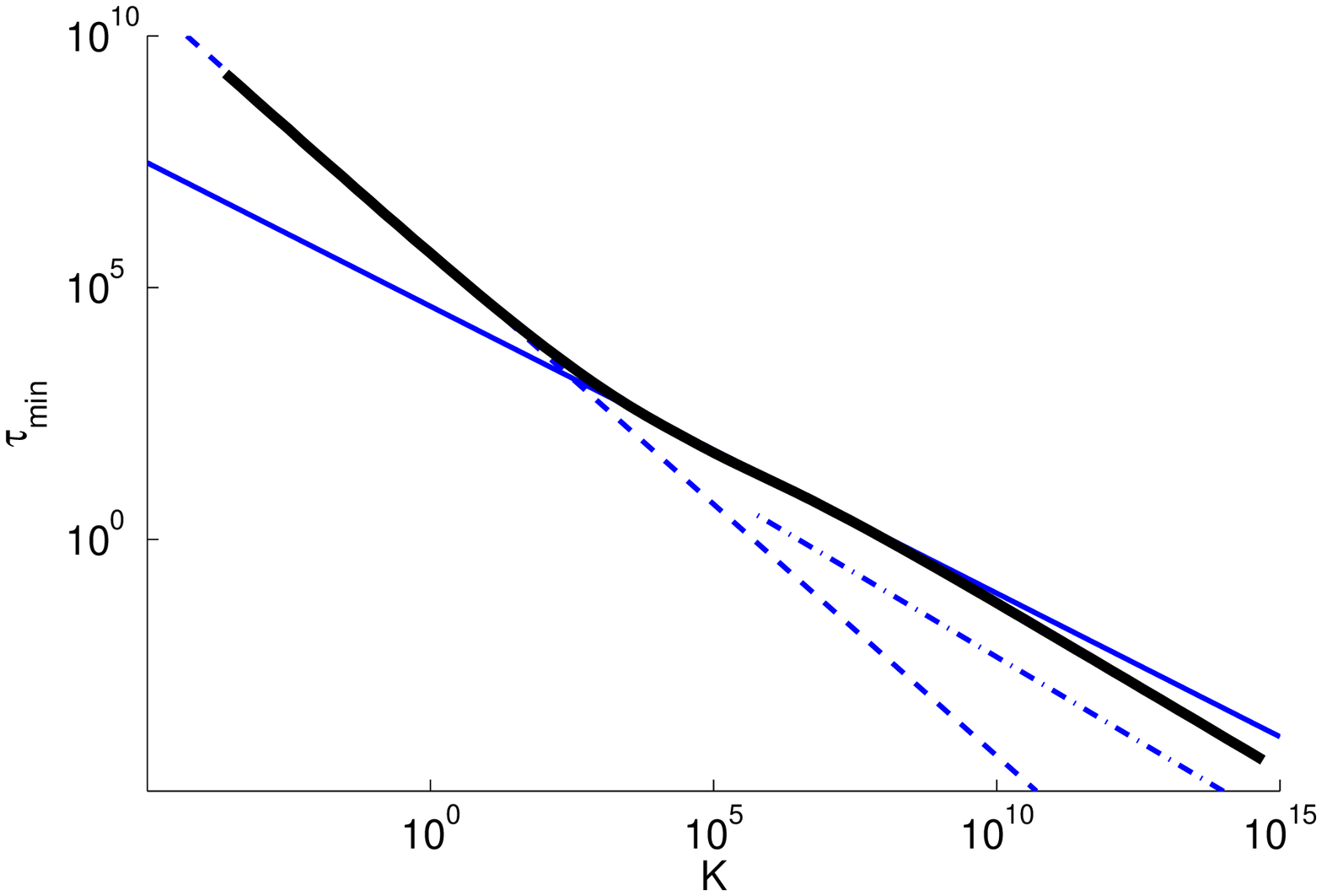}
\onefigure[scale=0.3]{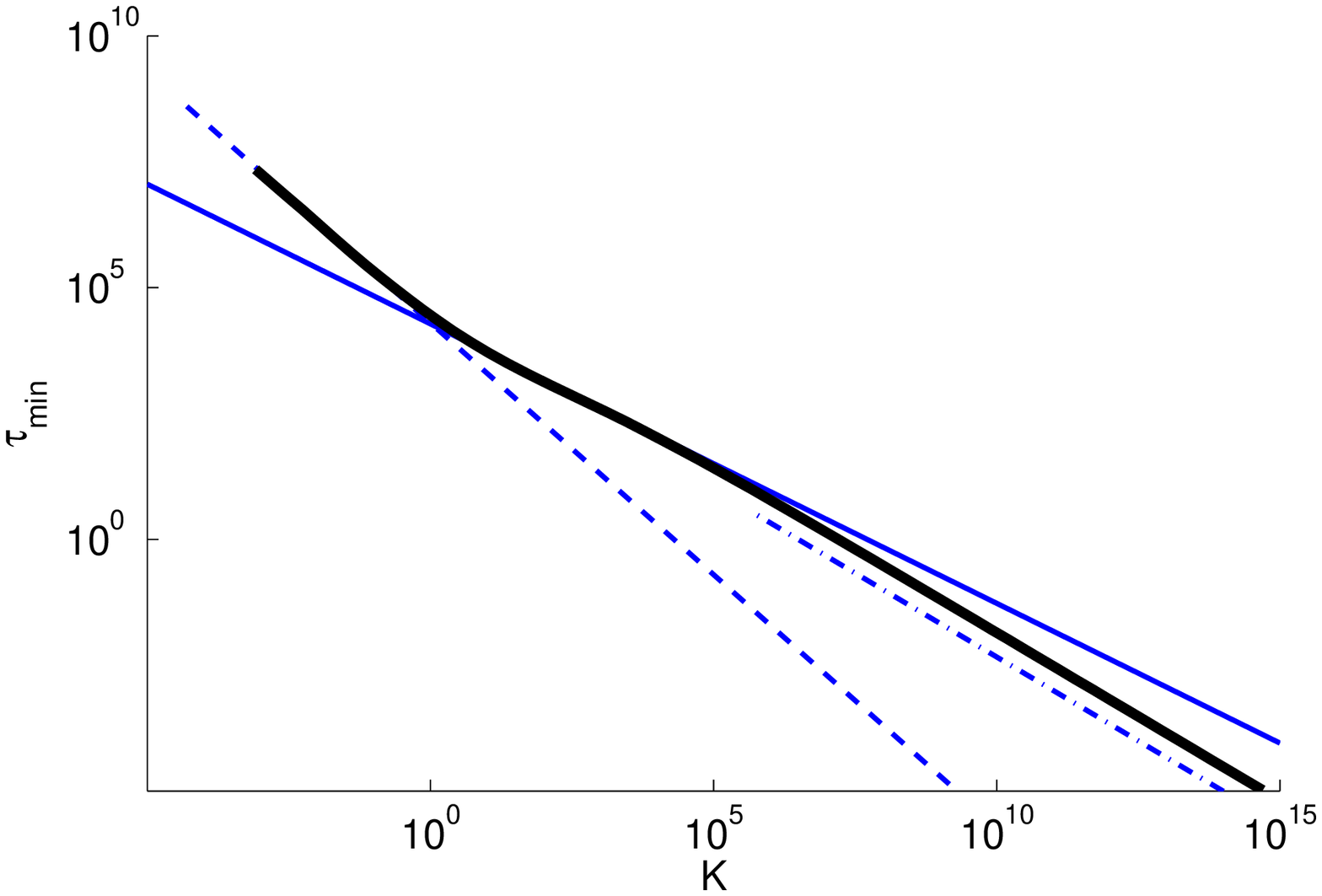}
\caption{First arrival time $\tau_{\min}$ vs.\ permeability $K$ 
for $\zeta=0.8$ (upper figure) and $\zeta=0.3$ (lower figure).  Each 
curve is based on one sample of size $512\times512$.  The slopes of
the straight lines are given in the main text.}
\label{fig:tauvsk}
\end{figure}

Since the first arrival time and the permeability are controlled by the same
aperture, it is possible to eliminate the aperture between them.  Hence, we
may express the first arrival time directly in terms of the permeability by
combining eqs.\ (\ref{1dpermlargeb}) -- (\ref{1arrival2d}). We show in 
Fig.\ \ref{fig:tauvsk} $\tau_{\min}$ vs.\ $K$ for two roughnesses, $\zeta=0.8$
and $\zeta=0.3$.  We expect that for small $K$, $\tau_{\min} \sim A/K^{1}$,
where $A$ is defined  in eq.\ (\ref{1arrival2d}).  For large $K$, we 
expect $\tau_{\min} \sim C/K^{2/3}$.  For intermediate $K$, we expect
$\tau_{\min} \sim A/K+C/K^{0.56}$ when $\zeta=0.8$ and $\tau_{\min} 
\sim A/K+C/K^{0.54}$ for $\zeta=0.3$.  
$C$ is defined in eq.\ (\ref{1arrival2d}).  
Straight lines with the appropriate slopes have been added in 
Fig.\ \ref{fig:tauvsk}.  For the intermediate region, we have used only
the term proportional to $C$.  

We have in this Letter shown that the permeability and the first arrival
time in dispersive processes are controlled by the same aperture length
scale in self-affine fractures.  We have also shown that the functional 
relation between the first arrival time and this aperture is linear,
see eq.\ (\ref{1arrival2d}).  The appropriate aperture is the minimal path
aperture defined in eqs.\ (\ref{bx}) and (\ref{bc}).  It is a generalization
of the  concept of the narrowest constriction that controls both the 
permeability and the first arrival time in one-dimensional fracture systems.
Whereas the scaling properties we report are specific to self-affine aperture
fields, the method of analyis based on optimal paths is not.

\acknowledgments
We thank J.\ P.\ Hulin for many interesting discussions.  A.\ H.\
thanks the Universit\'e de Paris-Sud 11 for financial support.
H.\ A. and L.\ T. thanks  the PICS ``The Physics of Geological Complex 
System'' and the R\'{e}seaux de Th\'{e}matiques de Recherches Avanc\'{e}es 
``Triangle de la Physique'' for financial support. 

\end{document}